\documentclass[pre,aps,10pt,twocolumn]{revtex4-1}
\usepackage{geometry}
\usepackage{graphicx}
\usepackage{amsfonts}
\usepackage{amsmath}
\usepackage{bm}
\usepackage{url}
\usepackage{color}
\usepackage{xfrac}
\usepackage{array,url,kantlipsum}

\newcommand{\ev}{{\bm e}}

\newcommand{\av}{{\bm a}}
\newcommand{\Ppara}{{\bm P}_\parallel}
\newcommand{\Pperp}{{\bm P}_\perp}
\newcommand{\W}{{\cal W}}
\newcommand{\T}{{\cal T}}
\newcommand{\Q}{{\cal Q}}

\newcommand{\overbar}[1]{\mkern 1.5mu\overline{\mkern-1.5mu#1\mkern-1.5mu}\mkern 1.5mu}

\begin{document}
\title{Local growth of icosahedral quasicrystalline tilings} 

\author{Connor Hann}
\affiliation{Physics Department, Duke University, Durham, NC 27708}

\author{Paul J.~Steinhardt}
\affiliation{Physics Department, Princeton University, Princeton, NJ 08544}

\author{Joshua E.~S.~Socolar}
\affiliation{Physics Department, Duke University, Durham, NC 27708}

\date{\today}

\begin{abstract}
Icosahedral quasicrystals (IQCs) with extremely high degrees of translational order have been produced in the laboratory and found in naturally occurring minerals, yet questions remain about how IQCs form. In particular, the fundamental question of how locally determined additions to a growing cluster can lead to the intricate long-range correlations in IQCs remains open. In answer to this question, we have developed an algorithm that is capable of producing a {\em perfectly} ordered IQC, yet relies exclusively on local rules for sequential, face-to-face addition of tiles to a cluster. When the algorithm is seeded with a special type of cluster containing a defect, we find that growth is forced to infinity with high probability and that the resultant IQC has a vanishing density of defects. The geometric features underlying this algorithm can inform analyses of experimental systems and numerical models that generate highly ordered quasicrystals.
\end{abstract}


\maketitle 
\section{Introduction}

Icosahedral quasicrystals (IQCs) with extremely high degrees of translational order have been produced in the lab \cite{TrebinBook} and found in naturally occurring minerals \cite{Bindi09}.  These materials possess icosahedral point group symmetry and quasiperiodic structure.  Their diffraction patterns consist of Bragg peaks at all integer linear combinations of a set of six independent basis vectors pointing to the vertices of a regular icosahedron, a dense set that includes wavevectors of arbitrarily small magnitude.  The presence of incommensurate collinear wavevectors gives rise to ``phason'' symmetries that have no analogue in crystals and strongly affect the elasticity and plasticity of the quasicrystal~\cite{Socolar86P}.

While the existence of IQCs is well established, the processes by which they form are not well understood.  
It is known that thermal annealing can improve the quality of a quasicrystal \cite{Calvayrac1990,Tsai1990}, but highly developed translational order has also been observed in rapidly quenched samples \cite{Tsai1990}, suggesting that nucleation and local growth kinetics produce a well-ordered IQC.  The kinetics of nucleation and growth from the liquid is also thought to play an important role in creating a sample that can be successfully annealed.  (See, for example, Refs.~\cite{Langsdorf1998,Schroers2000}.)  
There are, however, geometric features of quasicrystal structure and of defects associated with the phason degrees of freedom that raise questions about how any kinetic process can give rise to a well ordered sample.  

The atomic structure of a well ordered quasicrystal alloy can be described in terms of a space-filling tiling of two or more types of ``unit cells''~\cite{Socolar86,TrebinBook}.  If one imagines building the tiling one cell at a time, a difficulty is quickly encountered: the proper choice of which tile to add at some surface sites on the growing cluster can depend on choices that have been made in distant locations~\cite{Penrose1989}.  Growth of a perfect sample would appear to require interactions of arbitrarily long range, without which the growth process could not avoid the inclusion of a finite density of certain types of defects representative of phason fluctuations. The problem can be mitigated to some extent by allowing for annealing in a surface layer during the growth, but as long as the depth of the layer is finite, some degree of phason strain would appear to be inevitable.

In this paper we address the question of whether it is possible in principle for nucleation and growth to produce a {\em perfectly} ordered IQC.  We find that it is possible to produce with exceedingly high probability an IQC with a vanishing density of defects, using a local growth algorithm for sequentially adding tiles of two different shapes to a growing cluster.  By ``local,'' we mean that the choice of how to add a tile at any selected surface site is based only on information about the local environment at that site.  The infinite growth occurs when the algorithm is seeded with a special type of cluster containing a defect.  

The apparent requirement of nonlocality is avoided by introducing a distinction between {\em forced sites} and {\em unforced sites} on the surface of a growing cluster~\cite{Onoda88}.  At a forced site, the local configuration already present uniquely specifies how a tile (or cluster of atoms) can be added.  At an unforced site, there are at least two ways of adding tiles that would be consistent with the local environment, though possibly inconsistent with distant parts of the existing cluster.  To prevent inconsistent additions, the probability of adding any tile to a randomly selected surface site is taken to be zero at an unforced site and nonzero at a forced site.  In this way, information about distant parts of a cluster can be transmitted through locally forced additions until a tile is added near a previously unforced site that resolves any ambiguity, converting it to a forced one.  The question is whether, even in principle, a set of local forcing rules can be found that is sufficient to produce infinite growth rather than terminating with a cluster whose surface consists entirely of unforced sites. 

Our results are analogous to previously published results on the 2D Penrose tilings, which are quasicrystals with decagonal symmetry.~\cite{Penrose74,Onoda88,Ingersent90,Socolar91,Dotera91}  Important new features arise, however, due to the different topologies of 2D and 3D phason defects.  Unlike the 2D growth algorithm that produces a perfect Penrose tiling from a decapod seed that contains a single point defect~\cite{Onoda88}, the IQCs generated by our 3D algorithm necessarily contain line defects. The number of defects, however, grows only linearly with the cluster radius, leaving the bulk 3D sample with a vanishingly small density of defects, which occur only along special planes passing through the seed and correspond only to infinitesimal fluctuations of the phason field.

Recent numerical investigations~\cite{Engel2007,Engel2010,Achim2014,Steurer2015} and experiments~\cite{Edagawa2015} strongly suggest that favoring certain growth sites near the surface of a growing cluster can instill a high degree of long-range order.  It is not clear, however, how (or whether) these growth processes manage to avoid the generation of finite phason fluctuations or linear phason strain.  The present work shows that local growth can, in principle, account for the high degree of order in an IQC and elucidates mechanisms for generating nearly perfectly ordered, large samples via purely local growth kinetics.

In Sec.~\ref{sec:ammann}, we describe the tiling model due to Ammann that we use as the basis for our investigation.  Section~\ref{sec:growth}  
presents a local growth algorithm in which a tile is added to a surface vertex of a growing cluster in a manner determined completely by the already placed tiles that share that vertex.  Section~\ref{sec:dynamics} presents an analysis of the growth produced by the algorithm, showing that certain seeds give rise to nearly perfect growth that proceeds to infinity with a high probability.  We conclude with some remarks and discussion in Sec.~\ref{sec:discussion}.

\section{The Ammann Tilings} \label{sec:ammann} 
\begin{figure}
\centering
\includegraphics[width=0.45\textwidth]{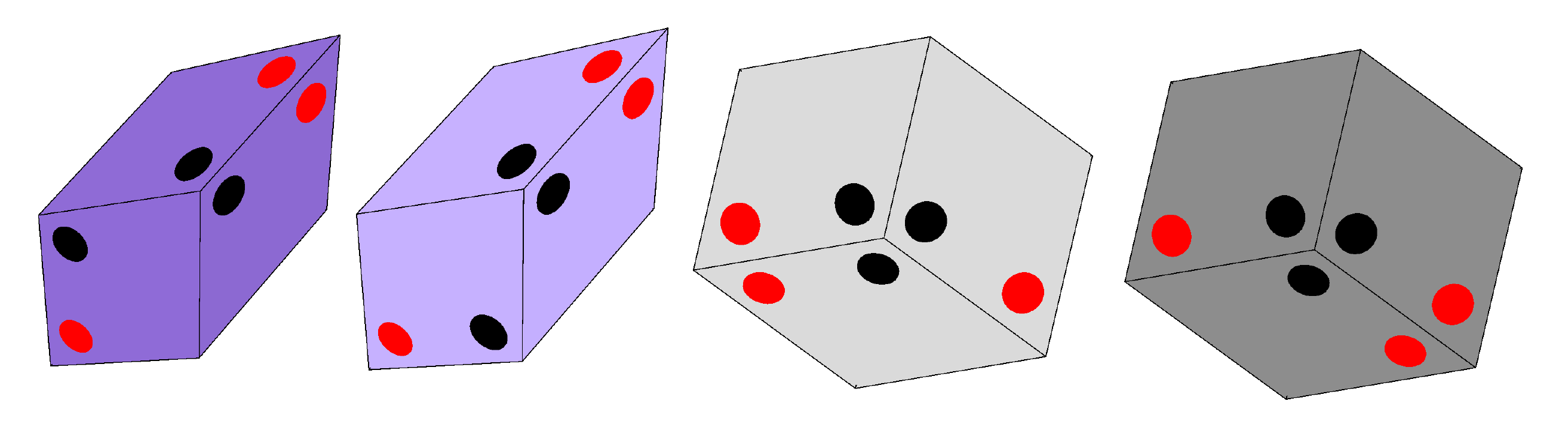}
\caption{Ammann tiles decorated with matching rule markings. The decoration of each tile is chiral, and both enantiomorphs are needed for each tile shape. Positions of dots on the faces not visible may be inferred from the visible dots: for the prolate tiles (left pair), the black dot on a hidden face is located in the same corner as the black dot on the corresponding parallel visible face. The red dot is located at the opposite corner from that of the corresponding parallel face.  The reverse is true for the oblate tiles (right pair).}
\label{proto}
\end{figure}

The tilings considered in this work are formed from oblate and prolate rhombohedra decorated as shown in Fig.~\ref{proto}.  Matching rules, which may be thought of as indicating energetically favored local configurations, specify that dots of the same color on a face shared by two tiles must coincide. 
These tiles and the rules enforced by the decorations were discovered by Ammann, and we refer to the class of defect-free tilings that can be made from them as {\em Ammann tilings}~\cite{Levine86,Socolar86}.  Ammann's markings of the rhombohedral tiles are known to be at least weak matching rules that enforce long range quasicrystalline order.~\cite{Levitov1988,Socolar90}  These particular rules have not been rigorously proven to be perfect matching rules (i.e., to force a single local isomorphism class of tilings), though closely related rules have been shown to do so.~\cite{Socolar86,Katz1989}  We proceed here on the assumption that the Ammann markings are indeed perfect matching rules, an assumption that is strongly supported by our finding that there exist rules for forcing growth of space-filling, infinite clusters.

The vertices of an Amman tiling may be obtained by direct projection of a subset of lattice points of a six-dimensional hyper-cubic lattice onto a three-dimensional subspace called the {\em tiling space} and denoted by $E_{||}$.  The projection onto the tiling space is defined as the projection that takes the six mutually perpendicular basis vectors of the hypercubic lattice into the six ``star vectors'' pointing to the vertices of an icosahedron.  
\begin{equation}
\label{eqn:star_vectors} 
{\bm e}_k = \begin{cases}
\frac{1}{\sqrt{5}}\left(  
2\cos\left( \frac{2 \pi k}{5} \right) ,
2\sin\left( \frac{2 \pi k}{5} \right) ,
1\right)  &k \leq 4\\
\left(0,0,1 \right)  &k = 5
\end{cases}.
\end{equation}
The projection of a hypercubic lattice point $\av = (a_0, a_1, a_2, a_3, a_4, a_5)$ onto the tiling space is 
\begin{equation}
\label{real_vert}
{\bm P}_\parallel(\av) = \sum_{k=0}^{5} a_k \ev_k\,.
\end{equation}

The subset of points that is projected is determined by a projection onto the orthogonal complement of the tiling space, generally referred to as ``perp-space" and denoted by $E_{\perp}$.  We define a set of perp-space star vectors:
\begin{equation}
\ev'_k = \begin{cases}
\ev_{\left\langle 3k \right\rangle } & k \leq 4 \\
-\ev_k  & k = 5
\end{cases}.
\end{equation}
The projection of $\av$ into $E_{\perp}$ is
\begin{equation}
\label{perp_vert}
\Pperp(\av) = \sum_{k=0}^{5} a_k \ev'_k \, .
\end{equation} 

To generate the vertices of an Ammann tiling, one defines a perp-space volume that is the projection by $\Pperp$ of a unit hypercube, which forms a rhombic triacontahedron called the ``perp-space window,'' designated $\W$.  The vertices of the tiling are the projections by $\Ppara$ of all hypercubic lattice points $\av$ for which $\Pperp(\av)$ lies within $\W$.  Note that the location of $\W$ in $E_{\perp}$ can be chosen arbitrarily, with different choices producing globally distinct Amman tilings that are locally isomorphic; i.e., that cannot be distinguished by examination of local configurations of any size.  Note also that $\W$ has the point group symmetry of a regular icosahedron.

Individual tiles may be constructed from the set of projected vertices by connecting each pair of vertices with unit separation. The above procedure yields two distinct tile shapes: one prolate rhombohedron, with edges parallel to $(\ev_0, \ev_1, \ev_5)$ or any symmetry related triple of star vectors; and one oblate rhombohedron, with edges parallel to $(\ev_0, \ev_1, \ev_2)$ or any symmetry related triple.

\section{Icosahedral Growth Algorithm} \label{sec:growth}

Using the above matching rules, and inspired by the results of Onoda et al.\ for the two-dimensional Penrose tilings \cite{Onoda88}, we consider a growth algorithm for Ammann tilings that relies exclusively on a local vertex rule to determine where and how additional tiles should be added. We first compile a catalog of all vertex configurations appearing in these tilings.  A complete specification of the catalog is presented in Table~1.

\begin{figure*}
\centering
\includegraphics[width=0.9\textwidth]{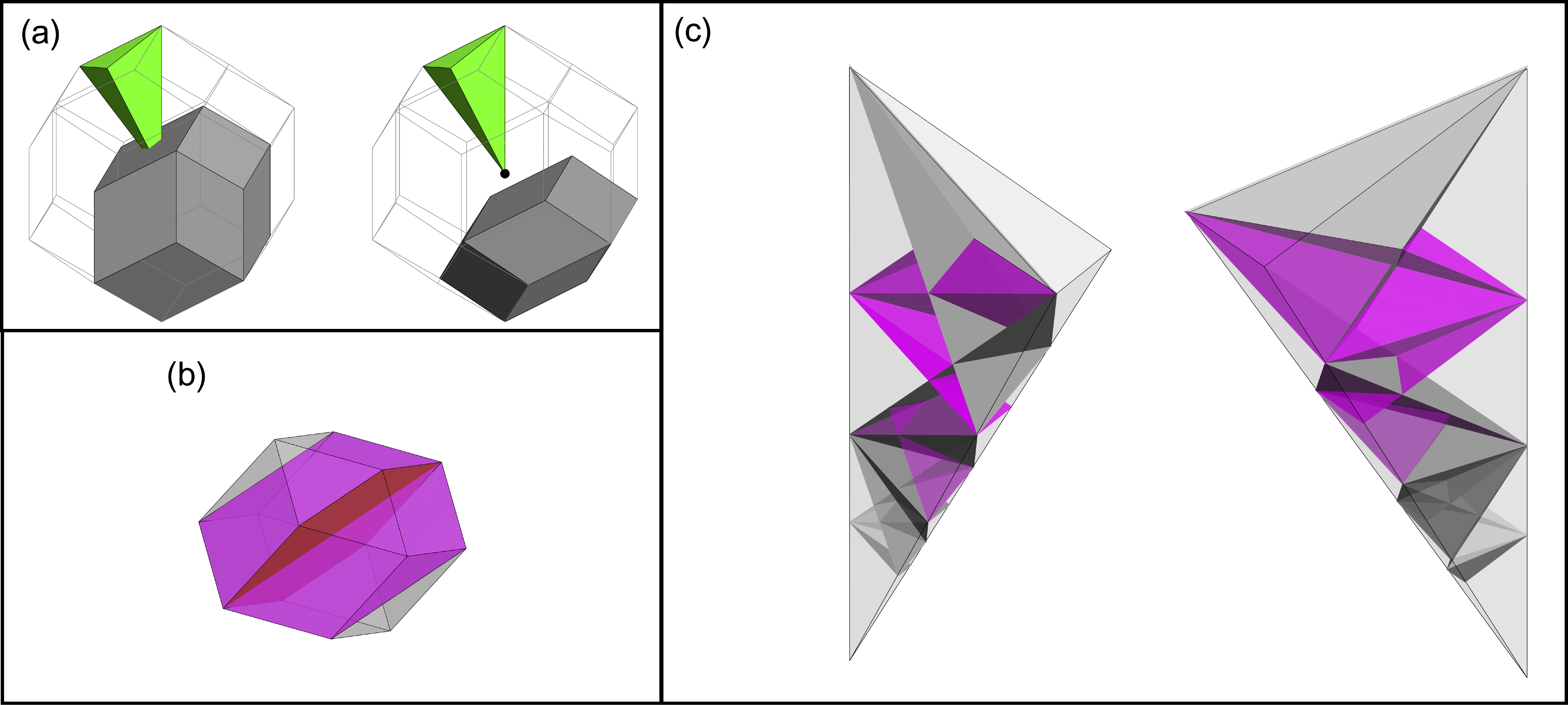}
\caption{Perp-space domains corresponding to distinct vertex configurations.  (a) The two types of vertex-face domains, with a fundamental domain of $\W$ shown in green.  (b) The division of a vertex-face domain into quadrants corresponding to distinct matching rule markings.  (c) Two views of the division of a fundamental domain into cells by the union of all icosahedral group operations on the vertex-face domain quadrants.   Colors are to aid the eye, with magenta planes corresponding to quadrant divisions and gray to vertex-face domain boundaries.}
\label{fig:facedomains}
\end{figure*}

To construct the catalog, we first identify the domains within $\W$ for which the corresponding vertex is part of a tile face with edges along two given star vectors $\pm\ev_i$ and $\pm\ev_j$.  There are two types of such vertex-face domains, depending on whether the angle between the two star vectors at the vertex is acute or obtuse.  The two types of domain are both rhombic dodecahedra \cite{Katz1989}, but are positioned differently within $\W$.  An example of each type is shown in Fig.~\ref{fig:facedomains}(a), which also shows a fundamental domain of $\W$ under the full icosahedral group $I_h$.  

The face corresponding to a given vertex-face domain may be decorated in any of four distinct ways by the matching rules markings; the red dot can be at either acute angle and the black at either obtuse angle.  These distinct markings correspond to distinct domains within the vertex-face domain, which gets divided symmetrically as shown in Fig.~\ref{fig:facedomains}(b).  (The dividing planes are determined by tracing possible paths of edges from the vertex until a vertex is placed that implies a tile specifying the location of the relevant mark.  See Katz  \cite{Katz1989} for a closely related analysis associated with a set of matching rules requiring 14 distinct decorations of the  prolate rhombohedron and 8 distinct decorations of the oblate one.)
Each quadrant of a given vertex-face domain corresponds to a distinctly oriented and marked face attached to a tiling vertex that projects into that domain in $\W$.
The number of distinct complete vertex configurations, up to $I_h$ symmetry operations, is obtained by examining a single fundamental domain of $\W$ to see how it is subdivided into cells by the boundaries of all of the quadrants of all of the vertex-face domains.  These boundaries, shown in Fig.~\ref{fig:facedomains}(c), form 39 cells.

\begin{figure}
\centering
\includegraphics[width=0.25\textwidth]{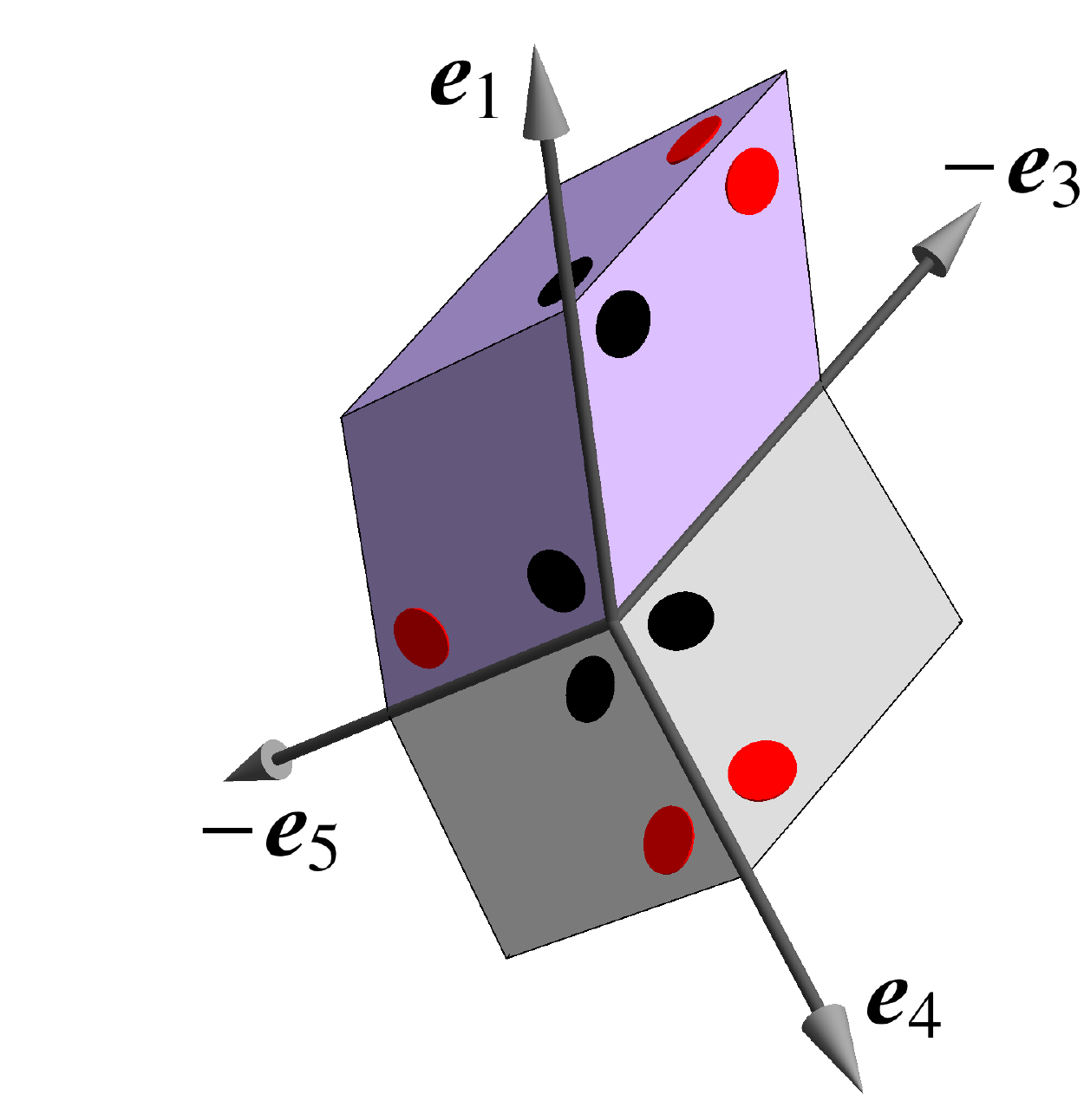}
\caption{The vertex configuration corresponding to the first row of Table~1.  See text for details.}
\label{fig:vert1}
\end{figure}

Each of the 39 cells corresponds to a unique vertex configuration specified by a row in Table~1.  An entry in the table specifies a particular face as follows.  The two numbers $ij$ specify that the edges of the face that emanate from the vertex of interest are $\ev_i$ and $\ev_j$, with $\overbar{x}$ indicating $-\ev_x$.  The order $ij$ indicates that there is a matching rule dot at the tip of edge $j$, and the arrow indicates the location of the other dot, with ``$\uparrow$'' indicating a dot near the vertex of interest and ``$\downarrow$'' indicating a dot at the opposite corner of the face.  Figure~\ref{fig:vert1} illustrates the meaning of the first row of the table.  Two of the tiles sharing the vertex are not shown so that we can see the vertex of interest.  Consider, for example, the face $\overbar{3}4\uparrow$.
It has a (red) dot at the corner along the $\ev_4$ direction, and a (black) dot at the vertex of interest.

\begin{figure}
\centering
\includegraphics[width=0.5\textwidth]{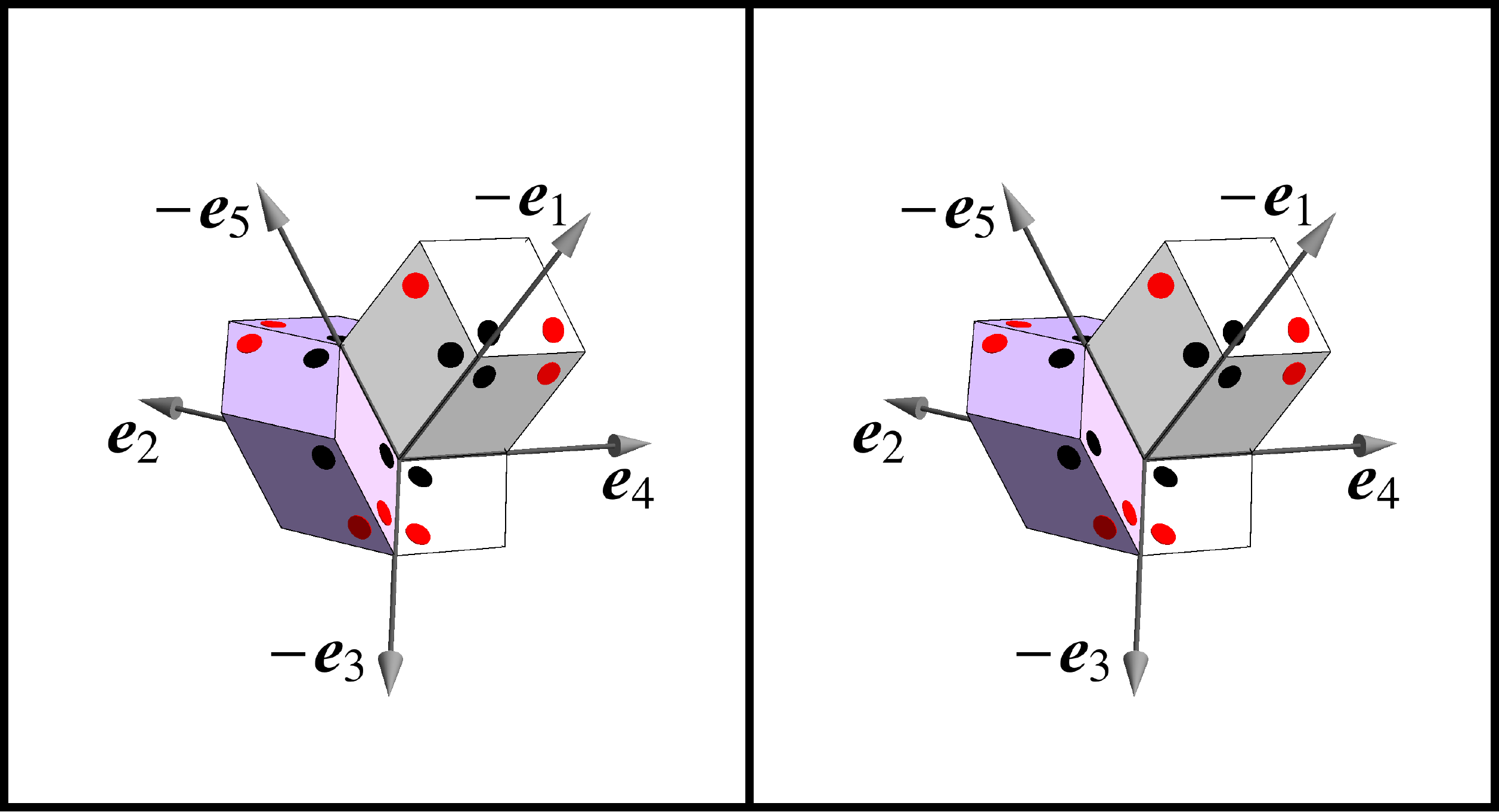}
\caption{Vertex configurations corresponding to the second and third rows of Table~1.  See text for details.}
\label{fig:vert2}
\end{figure}

Each colored band of rows in the table represents a set of cells that lie in the same set of vertex-face domains but not in the same quadrants of all of them; i.e., a set of cells specifying the same geometric vertex configuration but with different matching rule decorations.   As noted by Katz, there are 24 such cells~\cite{Katz1989}. Figure~\ref{fig:vert2} illustrates the difference between two rows in the gray band of three rows at the top of the table.  Again, two tiles have been removed to make the vertex visible.  The two vertices shown are identical except for the location of the black dot on the $\overbar{5}\overbar{3}$ face, which shows up in the table as a difference in the arrow directions for the first two rows in the band.

The existence of 24 distinct vertex configuration geometries and 39 distinct configurations when matching rules are included has been confirmed by direct computer assisted inspection of regions of Ammann tilings with tens of thousands of tiles.

Tiles are added to a growing cluster only at sites where the choice of what to add is uniquely determined by the requirement of consistency with the vertex catalog.  We identify which, if any, of the vertices in the catalog represent possible ways of completing a given vertex.  Any tiles that are present in all of the possible complete configurations and not already present in the cluster are labeled {\em forced tiles}.  At each time step, a forced tile is selected at random and added to the cluster.  The procedure is repeated until there are no forced tiles at any vertex on the surface of the cluster. 

In more precise terms, the algorithm may be described as follows:
Let $\Q_w$ be the set of oriented tiles comprising the vertex $w$ in the catalog.  Let $\T(v)$ be the set of tiles that intersect at a vertex $v$ and have already been placed in a growing cluster.  If $\T(v)$ is a subset of $\Q_w$, then let $\T_w(v)$ be the complement of $\T(v)$ in $\Q_w$; i.e., $\T_w(v)$ is the set of tiles that must be added to $\T(v)$ to complete the vertex $w$.

A vertex in a growing cluster is called {\em complete} if it is fully surrounded by tiles.  In almost all cases, a complete vertex will have $\T(v) = \Q_w$ for some $w$.  Complete vertices for which $\T(v)$ is {\em not} in the allowed vertex catalog are {\em defects}.

Incomplete vertices may be {\em forced} or {\em unforced}.   Consider all of the sets $\Q_w$ associated with catalog vertices that contain $\T(v)$ as a subset, and let $\T_f(v)$ be the intersection of all of those $\Q_w$'s.  If $\T_f(v)-\T(v)$ is not empty, then the vertex $v$ is forced, as there is at least one unplaced tile in $\T_f(v)$ that exists in all possible completions of $v$.  The tiles in $\T_f(v)$ are called {\em forced tiles}.  If $\T_f(v)$ is the empty set, then $v$ is an unforced vertex, meaning that there are two or more ways to complete the vertex that do not share any tiles that have not already been placed.

The growth proceeds by the sequential addition of forced tiles.  When a tile is added, new forced vertices may be created, and the growth continues until no forced vertices remain.  As long as there are no defects in the cluster, the order in which forced tiles are added makes no difference.  Small differences (discussed in detail below) can arise when the cluster contains defects.  In the present work, the order of additions is random: at each step a vertex is selected at random from the current set of all forced vertices, all forced tiles at that vertex are added, and the list of forced vertices is updated.

Note that the growth procedure does not rely on any global information about the position of vertices within the tiling, nor does it rely on information about the positions or orientations of any tiles beyond those that share a vertex with the added tile.  In physical terms, the information about distant structures in the growing cluster is tracked only through the requirement that no tiles be added to unforced sites, and this requires only local information at each surface site. 

\section{Growth dynamics} \label{sec:dynamics}
\subsection{Worm planes}
\begin{figure}
\centering
\includegraphics[width=0.5\textwidth]{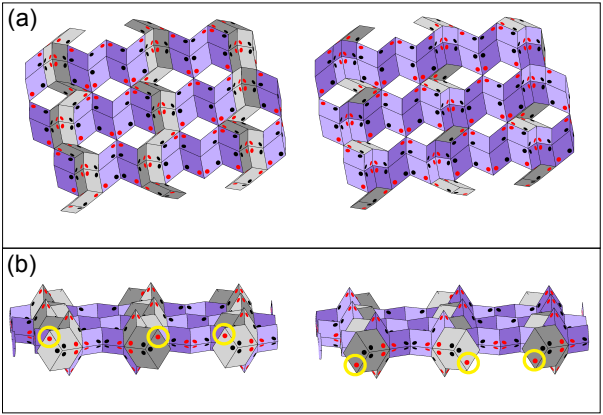}
\caption{a) Worm planes of opposite orientations. These two planes are composed of different tiles, yet have the same matching rule markings on their top and bottom surfaces. Prolate tiles are dark and light purple; oblate tiles are dark and light gray.  White rhombi are tile faces that are not part of any tile in the worm plane.  (b) Matching rule markings on the boundary of a worm plane that specify its orientation. }
\label{worm_plane}
\end{figure}

A key to understanding the growth process generated by the above algorithm is the structure we call a {\em worm plane}, which is analogous to a linear worm in the Penrose tilings~\cite{Gardner77,Socolar86P}.   A portion of a worm plane is shown in Fig.~\ref{worm_plane}(a).  The crucial feature of this planar slab of tiles is that the vertices in the interior of the slab can be moved vertically so as to create a second version of the slab that has exactly the same outer surfaces, including the matching rule markings, as the original, while the markings on interior faces in two slabs differ, as indicated in Fig.~\ref{worm_plane}(b).   The operation that moves moves all of the interior vertices and changes all markings on the interior faces accordingly is called a {\em worm flip}.

If a portion of the surface of a cluster corresponds to the surface of a worm plane that has not yet been placed, it will contain no forced vertices.  The worm can be added in either of its two possible orientations, thus there are two distinct ways to complete any given surface vertex.  Once a choice is made for one vertex on the worm plane surface, all of the others will be forced.

Worm planes are important structural elements for two reasons.  First, the choice of orientation of a given worm plane must be coordinated within the worm plane itself.  If different choices are made for the orientation of the worm in two half-planes, a line of defects will necessarily be created where the two halves of the worm are joined.  The growth algorithm avoids such defects by filling forced vertices first.  Once the orientation of a worm plane is determined at a single vertex, the rest of the worm plane will be filled due to forced additions that propagate the information about the worm orientation to the full plane.

Second, the orientations of parallel worm planes must be correlated in subtle ways.  In certain configurations, the necessary orientation of a worm plane can be determined by the orientation of a parallel worm plane that is far away.  If arbitrary choices were made for the two orientations, the subsequent addition of forced tiles might eventually lead to conflicting choices for the orientation of a worm plane transverse to the first two, thereby generating a line of defects somewhere between the two original worm planes.

Our growth algorithm avoids this second problem by simply halting when there are no forced vertices on the surface of the growing cluster.  This occurs when the surface consists entirely of worm planes oriented such that no forced vertices occur along edges or at the corners of the faceted cluster.  
(We have characterized the possible dihedral angles and solid angles at the corners that have no forced vertices, but we omit the details here as they are not relevant to the main results.)
To avoid this type of arrest in the growth, we introduce special seeds that nucleate infinite growth as described below.

A perfect, infinite Amman tiling contains worm plane regions with 15 different possible normal vectors, corresponding to the planes of mirror symmetry of the icosahedron.  Typical worm plane regions are bounded by intersecting worm planes with different normal vectors.    
At these intersections, the orientation of one worm plane can force the orientation of the other.
There may be as many as four intersecting infinite worm planes in the tiling.

\subsection{Seeds for growth: Triacontapods}

Consider a finite, closed surface comprised of marked rhombic faces.  If the surface can be found within a perfect Ammann tiling, we refer to it as the surface of a ``legal'' cluster of tiles.  If it cannot, we call it an ``illegal surface.''
Given any legal cluster as a seed, growth through the addition of forced tiles must eventually halt.  To see why, consider the structure of the finite cluster in perp-space.  Recall that distinct positions of the window within $E_{\perp}$ specify distinct tilings.    Thus the vertices of an infinite tiling are uniquely determined only when the location of the window is fixed.  We know, however, that any finite portion of an Ammann tiling can be found (infinitely many times) in any Ammann tiling, which means that the finite cluster cannot precisely fix the location of the window.  By definition, forced growth cannot rule out any of the possible windows that contain the points in the original cluster.  In other words, forced growth can never result in a perp-space point being placed outside the the hull defined as the intersection of all windows $\W$ that contain the points that have already been placed.  Growth of an infinite tiling, however, must produce perp-space points that fill an entire window.  Thus forced growth from a legal seed cannot yield an infinite tiling.

In order for forced growth to proceed indefinitely, we must begin with an illegal seed containing a defect that determines the precise location of the window.   Such a seed can be constructed via analogy to the decapod seeds that generate infinite forced growth in the Penrose tilings.\cite{Onoda88,Ingersent90,Socolar91,Dotera91}
For the Ammann tilings, a suitable seed is a {\em triacontapod}, a rhombic triacontahedron with exterior matching rule markings.  An example is shown in Fig.~\ref{legalseed}.

\subsection{Legal seeds}
\begin{figure}
\centering
\includegraphics[width=0.15\textwidth]{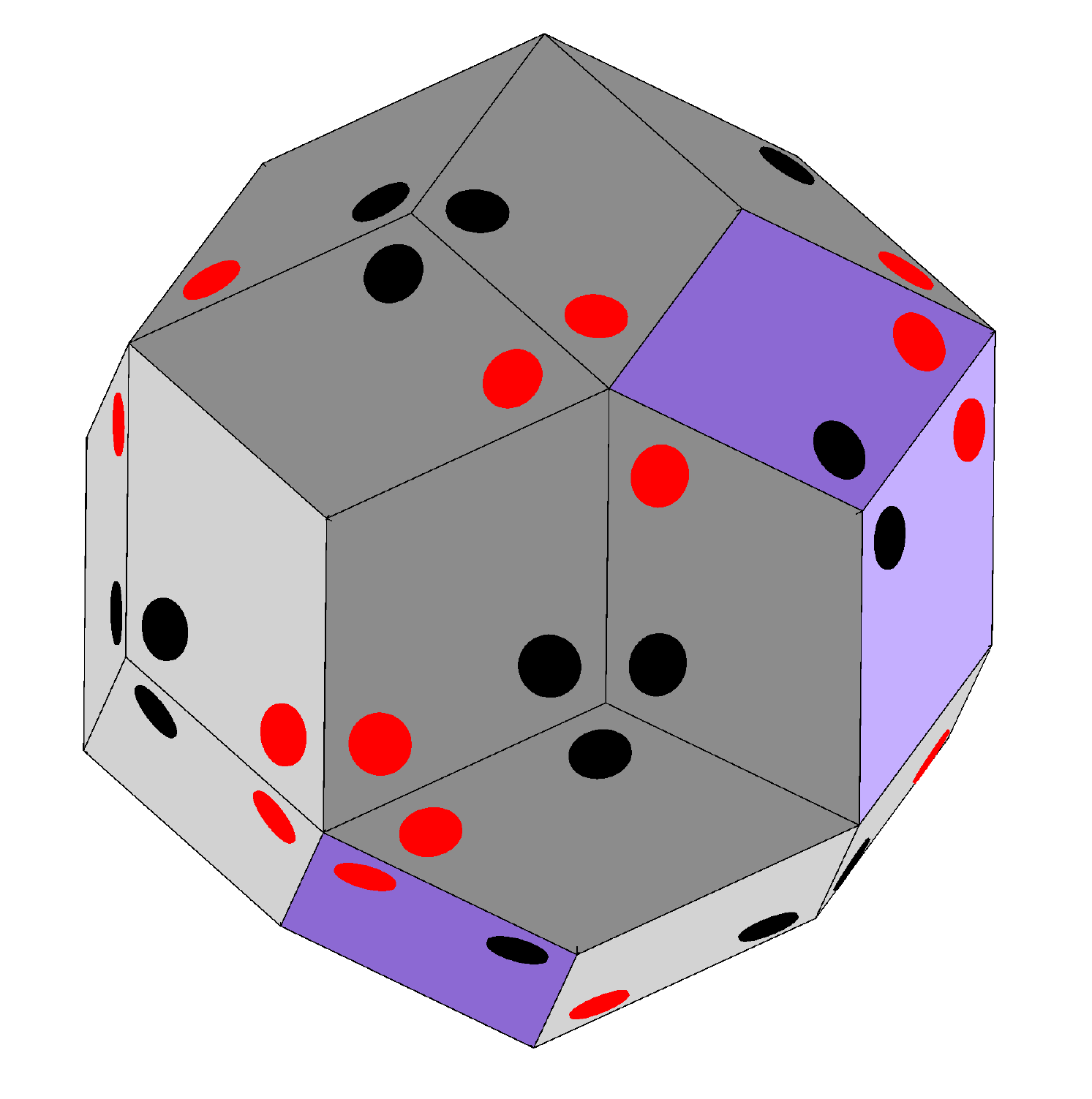}
\includegraphics[width=0.3\textwidth]{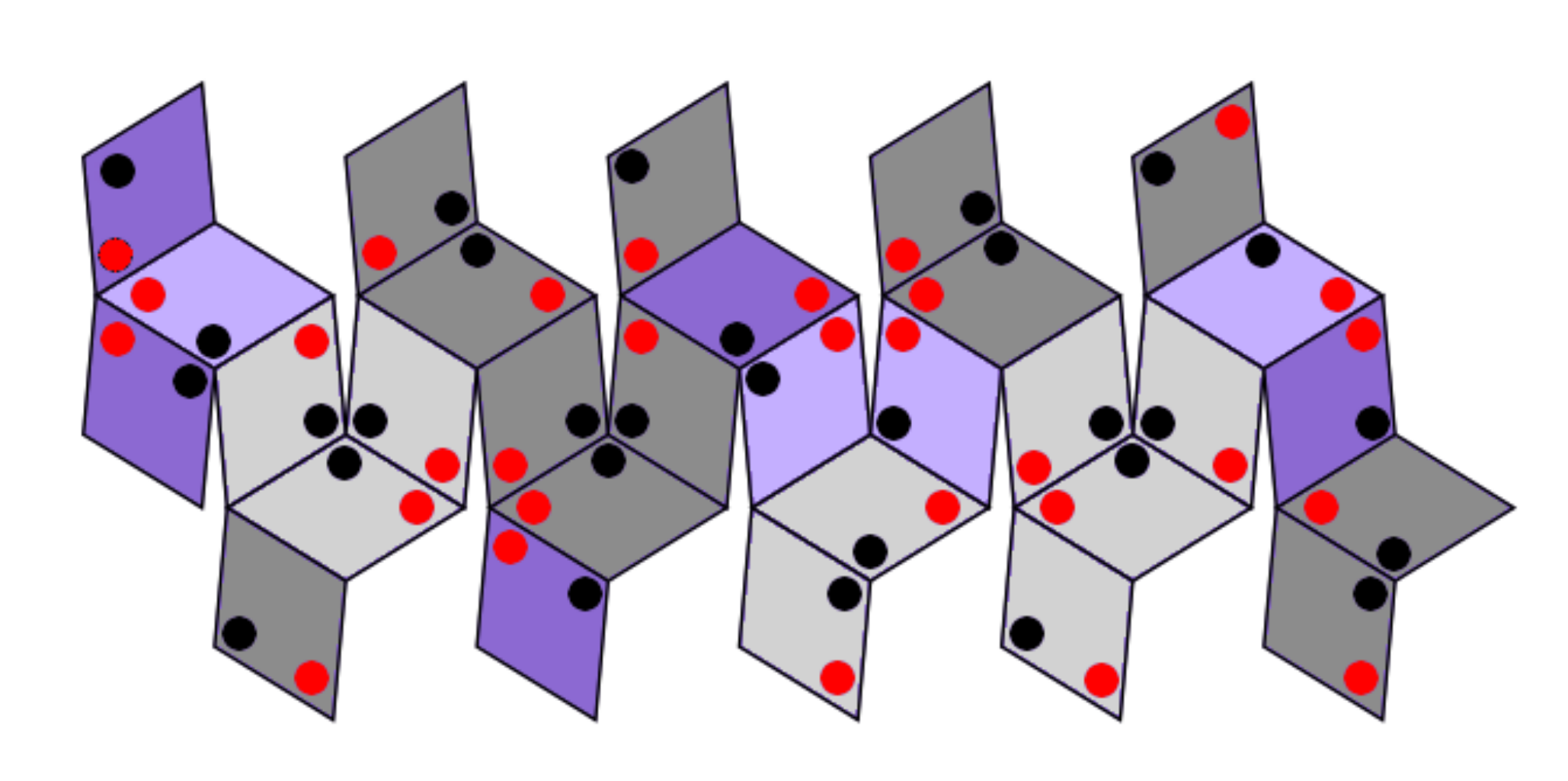}
\caption{A legal triacontapod and its unfolded net of faces. There exists an Ammann tiling in which 15 infinite worm planes intersect at this triacontapod.}
\label{legalseed}
\end{figure}

\begin{figure}
\centering
\includegraphics[width=0.5\textwidth]{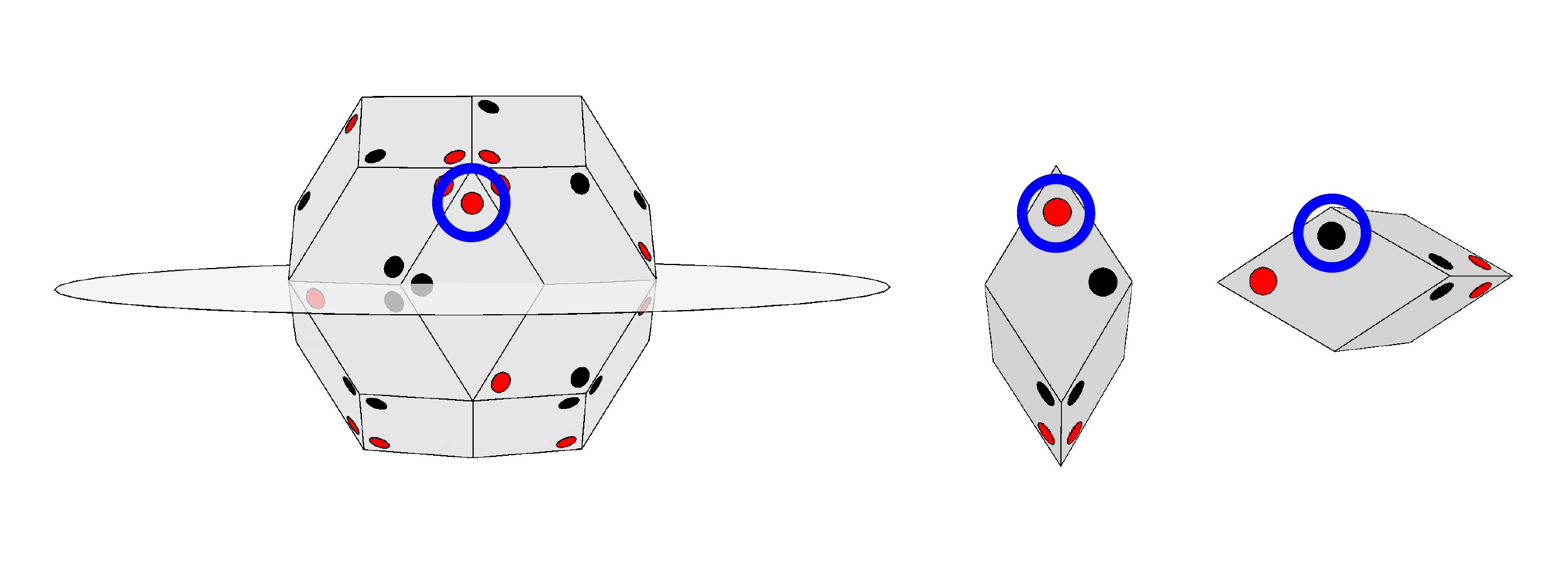}
\caption{Triacontapod seeds dictate the orientation of worm planes. A seed is shown with a worm plane represented symbolically by the horizontal gray plane. The matching rule dot circled in blue dictates the orientation of the worm plane. To satisfy the matching rules, tiles in the interior of the worm plane must be oriented such that the dot (red or black) on each face perpendicular to the plane lies on the same side of it.}
\label{seed_dictate_worm}
\end{figure}

The markings on the triacontapod of Fig.~\ref{legalseed} are consistent in the sense that this configuration does appear in the Ammann tilings, and the triacontahedron can be filled in with tiles that obey the matching rules everywhere.  There is exactly one Ammann tiling that has 15 infinite worm planes all intersecting to form this triacontapod.  Four of these pass through the triacontapod to form perfect worm planes; the others are disrupted in the interior of the triacontapod but are otherwise perfect.  The orientation of each worm plane is dictated by the markings on two opposite faces of the triacontapod, as illustrated in Fig.~\ref{seed_dictate_worm}.  In more general cases (i.e., triacontapod defects) we will assign dots on the seed's surface manually without worrying about whether the interior of the seed can be consistently tiled.  Thus we drop the color distinction between the four types of tiles when showing a triacontapod.
There is only one legal triacontapod, up to symmetry operations on the icosahedron, shown in Fig.~\ref{legalseed}.   Any other pattern of marks on the triacontapod makes for an illegal seed.

\begin{figure}
\centering
\includegraphics[width=0.3\textwidth]{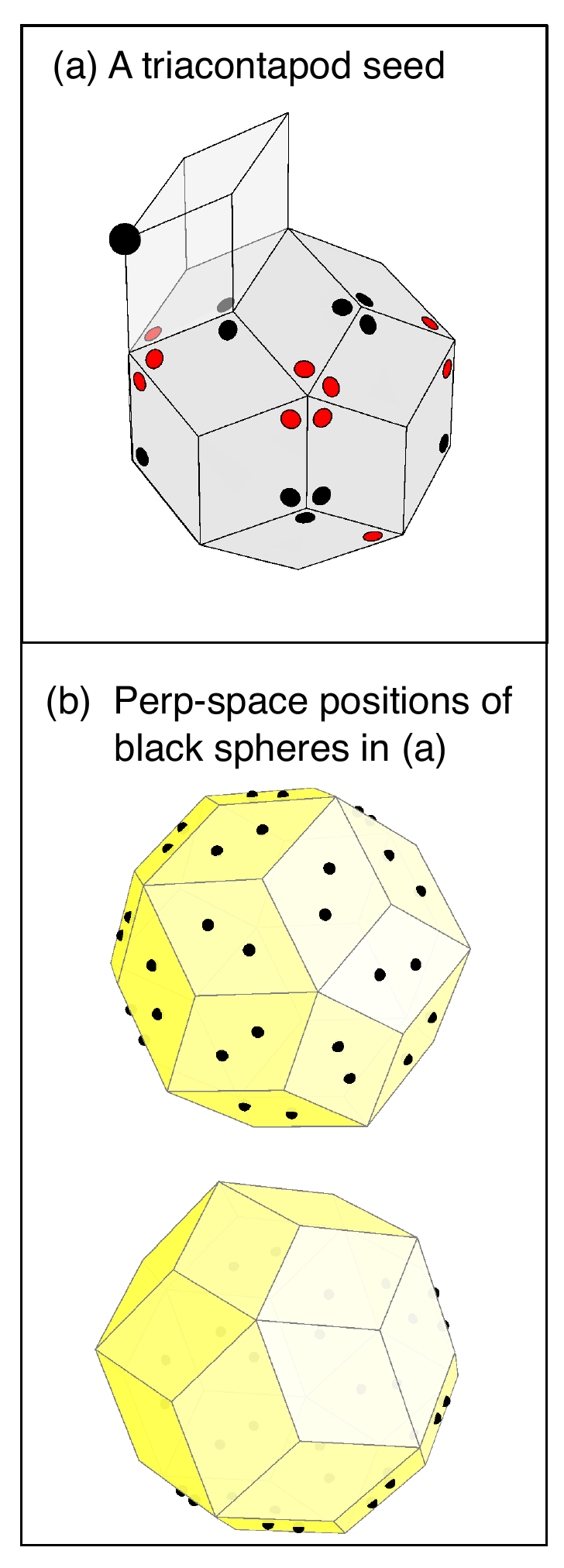}
\caption{(a) A triacontapod with a prolate tile attached.  The orientation of the tile is dictated by the covered red dot on the triacontapod surface.  The black sphere marks the vertex on the prolate tile that lies in the interior of a worm plane.  (b)  Two views of a perp-space window and the 30 projected vertices (black spheres from (a)) for a legal triacontapod seed. The perp-space window, shown as a transparent yellow triacontahedron (not to be confused with the real space triacontapod), is displayed from two opposite perspectives.  The 30 vertices lie in one hemisphere of the window.}
\label{legal_window_gray}
\end{figure}

When we speak of using a triacontapod as a seed for growth, we assume that the seed includes prolate tiles covering all of the faces of the triacontapod.  One such tile is shown in Fig.~\ref{legal_window_gray}(a).
The red dots on a triacontapod determine the orientations of these prolate tiles and hence the positions of 30 vertices like the one marked by a black sphere in Fig.~\ref{legal_window_gray}(a), whose normal projection onto the triacontapod face in question lies within that face.  Each of these vertices lies in the interior of a worm plane, determining its orientation.  It is instructive to examine the locations of these 30 vertices in perp-space.   Figure~\ref{legal_window_gray}(b) shows their locations for the case of a legal triacontapod, obtained from Eq.~(\ref{perp_vert}) using indices taken from Eq.~(\ref{real_vert}). The figure shows one possible perp-space window containing those points (not to be confused with the real space triacontapod!).  For the window shown, the points lie precisely on exterior facets, and for the tiling determined by this window, the triacontapod lies at the intersection of 15 infinite worm planes.  As must be the case for a finite legal seed, however, there exist other windows that contain all of the points.  Roughly speaking, the points all lie in one hemisphere of the window shown, and the window can be shifted in the direction of the pole of that hemisphere and still contain all 30 points.

\subsection{Illegal seeds}

In order to force growth to infinity, a seed must fully constrain the position of the perp-space window.~\cite{Dotera91}   We can arrange for a triacontapod to uniquely determine the window by choosing the markings such that the black dots of Fig.~\ref{legal_window_gray}(b) fall on facets that do not all lie in any single hemisphere.  This can be accomplished, for example, by moving the red mark that specifies the orientation of the prolate tile in Fig.~\ref{legal_window_gray}(a) to the opposite corner of the face it lies on.  The resulting flip of the tile causes the black dot in perp-space to jump to the opposite face of the window.

\begin{figure}
\centering
\includegraphics[width=0.50\textwidth]{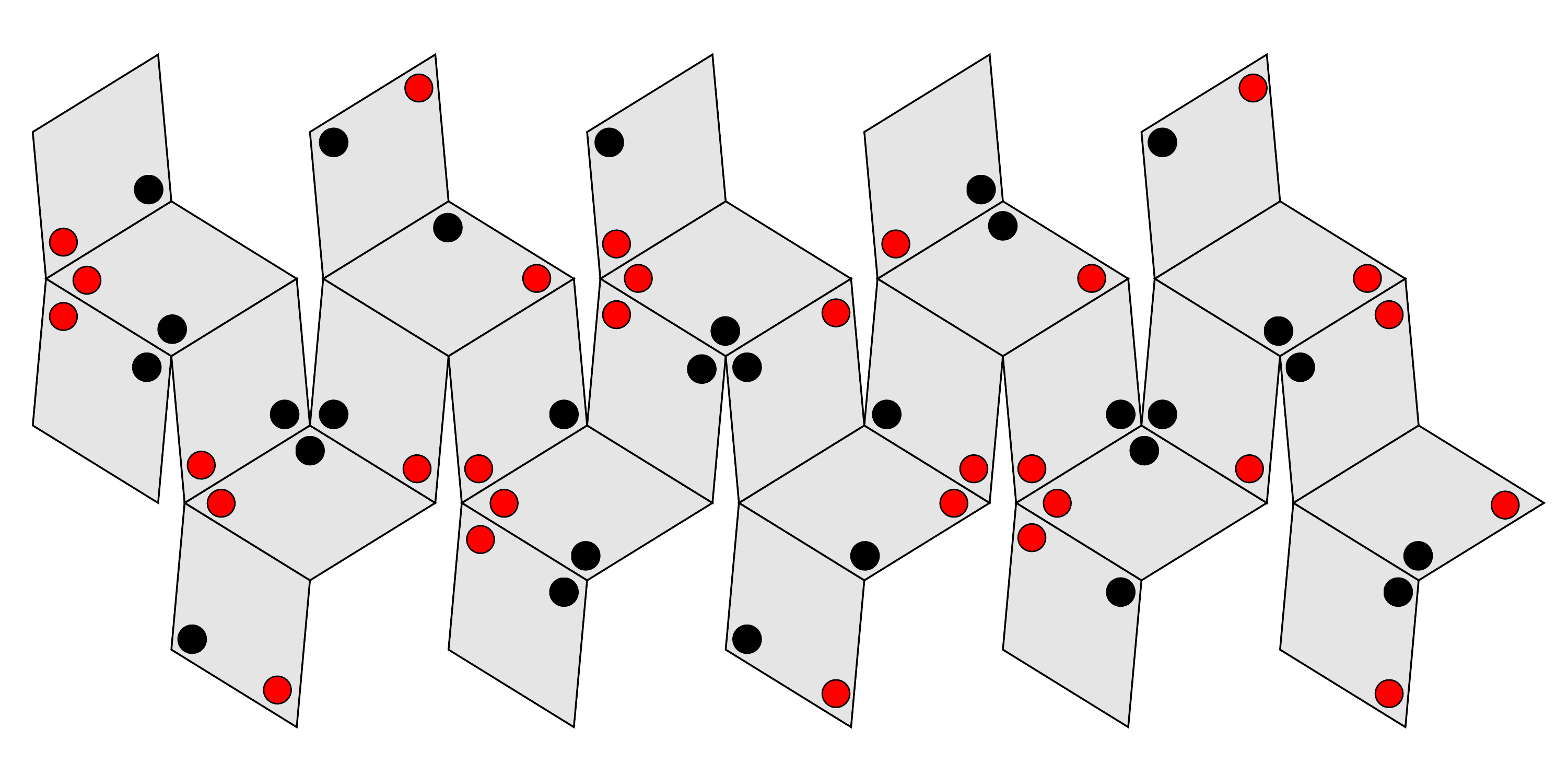}
\caption{An unfolded net representation of an illegal seed that constrains the perp-space window.  (Compare to Fig.~\ref{legalseed}.) }
\label{net_infinite}
\end{figure}

\begin{figure}
\centering
\includegraphics[width=0.25\textwidth]{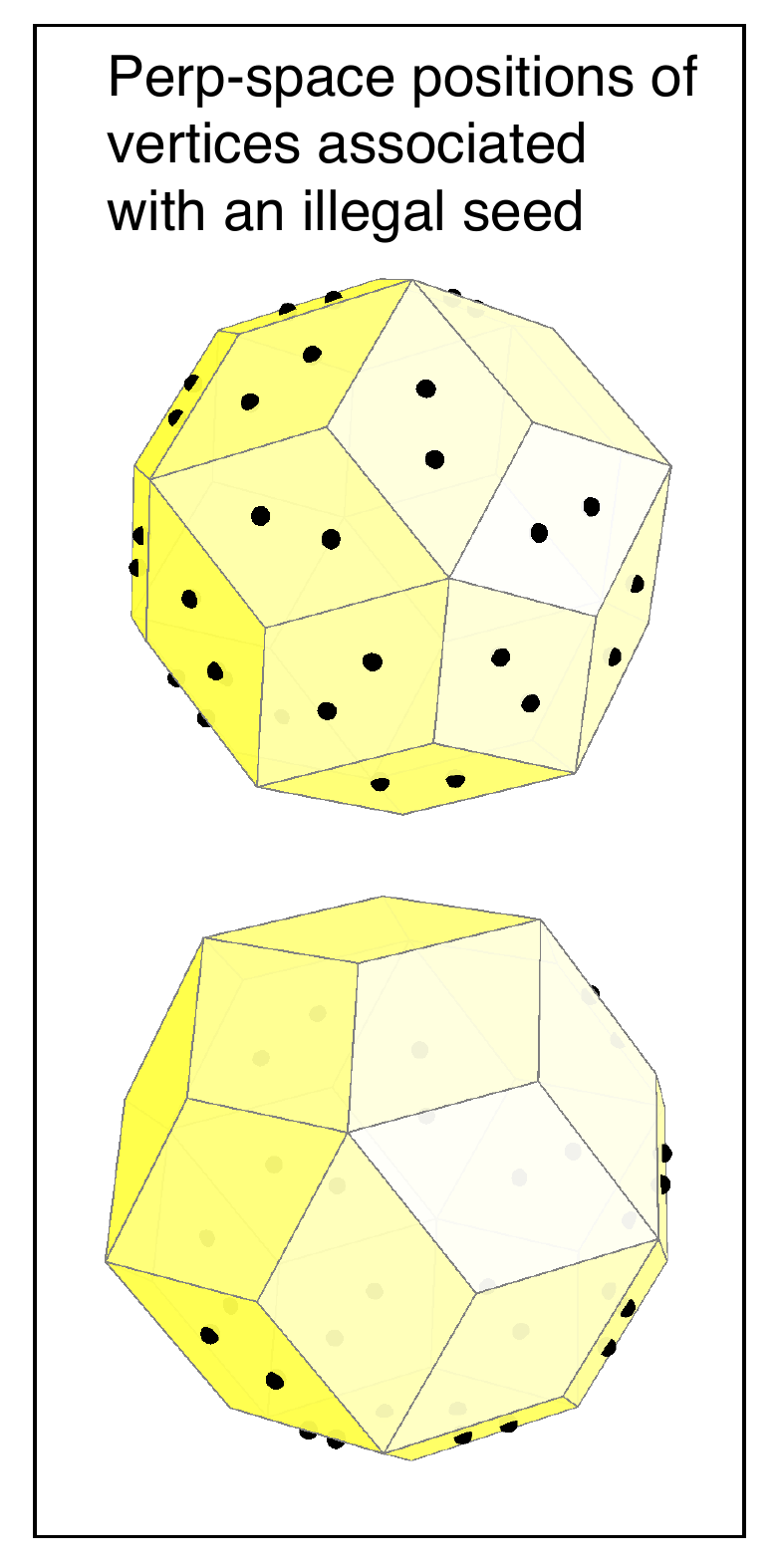}
\caption{Perp-space positions of projected vertices (black spheres of Fig.~\ref{legal_window_gray}(a) for an illegal triacontapod seed that fully constrains the perp-space window.  The difference between this figure and Fig.~\ref{legal_window_gray}(b) is that two vertices have been moved to the opposite side of the window.}
\label{constrained_window_gray}
\end{figure}

An example of an illegal seed is shown in Figure~\ref{net_infinite}, and a plot of the forced vertices in perp-space for the same seed is shown in Figure~\ref{constrained_window_gray}.  The location of the perp-space window is fixed; attempting to shift the window in any direction will move at least one vertex outside of the window.  This implies that there is at most one tiling that is consistent with the matching rules everywhere outside the seed, and so it is possible, but not guaranteed, that the seed forces growth to infinity. 

For the 2D Penrose tilings, Onoda et al.\ pointed out that there exist tilings that obey the matching rules everywhere outside an illegal decapod and that, for some illegal decapods, the surface of any cluster containing the decapod must always have at least one forced vertex.  (See also \cite{Ingersent90,Socolar91,Dotera91}.)  For such decapods, the sequential addition of forced tiles never halts and never produces a matching rule violation.

\begin{figure}
\centering
\includegraphics[width=0.25\textwidth]{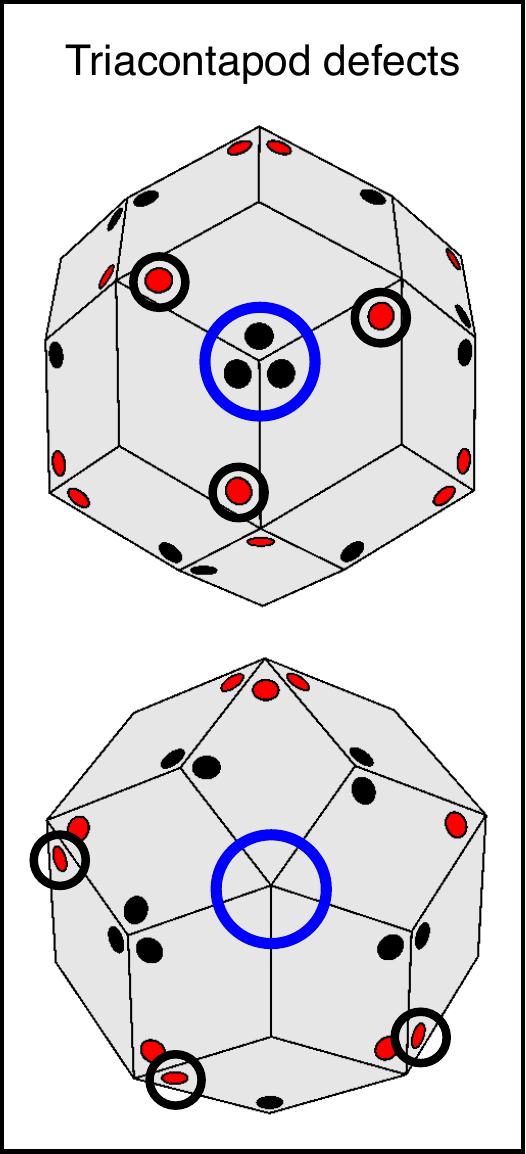}
\caption{The two varieties of triacontapod vertex configurations (circled) that force defects to appear during growth.  Every illegal triacontapod contains at least two vertices in this class.}
\label{defective_vertices_gray}
\end{figure}

For the 3D Ammann tilings, the situation is different: any tiling that contains an illegal triacontapod must contain matching rule violations outside the seed. To see this, consider the vertices of the illegal seed shown in Figure~\ref{defective_vertices_gray}. Notice that the red dots, circled in black, are three-fold symmetric about the vertex circled in blue. Such a configuration does not appear in any vertex in the catalog.  Similarly, for the lower image, it can be determined by inspection that while the vertex circled in blue can be completed without any matching rule violations, the dots circled in black will force the creation of a vertex that cannot be legally completed. In general, growth from any seed with either a three-fold symmetric vertex, as illustrated in the upper image, or a vertex with a chiral pattern of dots as illustrated in the lower image, must produce additional defects.  An exhaustive search through all illegal seeds reveals that each possesses at least two such vertices. 

Though the creation of defects during the growth might be expected to prevent forced growth from proceeding, it turns out that the algorithm can and does accommodate these defects and still generates a space-filling tiling by adding only to forced vertices as originally defined.  The tiles surrounding the illegal vertices get added due to forcing from other nearby vertices.  A matching rule violation occurs on a single face shared by two tiles that get incorporated into the bulk as growth proceeds.  The precise location of the mismatch may depend on the order in which forced tiles are added, but it must occur somewhere along the row of tiles that share faces parallel to the mismatched face.

Moreover, such defects do not disrupt the overall quasiperiodic order.  It remains true that forced growth can never produce a vertex that lies outside the perp-space hull determined by the tiles that have already been placed.  This means that the only defects in the tiling occur outside the triacontapod seed lie within the infinite worm planes, whose interior vertices lie on the boundaries of the perp-space window.  These defects must manifest as vertices that lie on opposite facets of the window.  The bulk of the tiling is therefore defect free, and as the inconsistent worm plane orientations on two halves of an infinite worm plane meet along a line of defects, the number of defects is expected to grow only linearly with cluster radius and therefore have a vanishingly small density in the infinite tiling.

\begin{figure}
\centering
\includegraphics[width=0.4\textwidth]{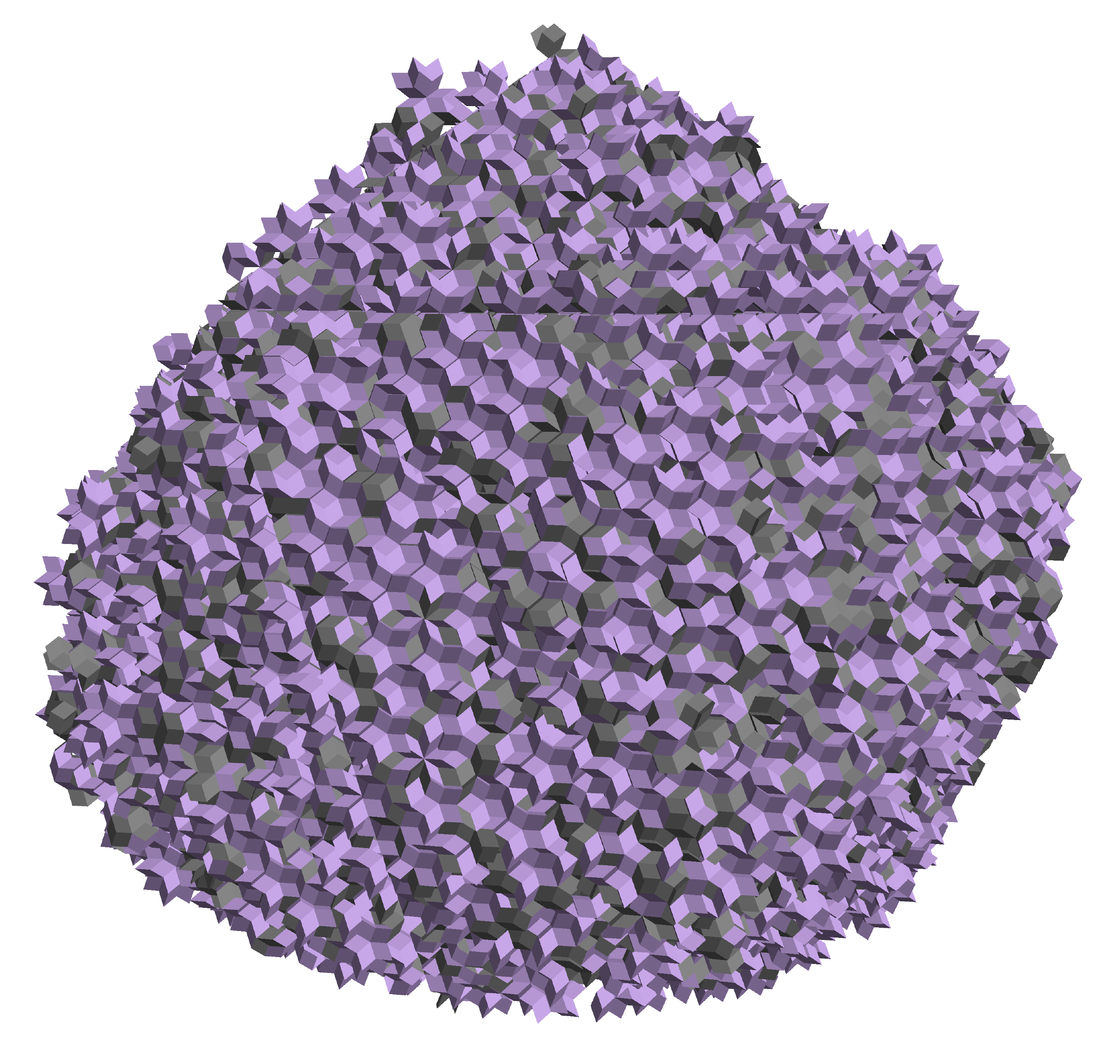}
\caption{A tiling grown from the seed of Figure~\ref{net_infinite}. This cluster contains approximately 100,000 tiles. Larger clusters have been grown from this seed, and none has yet encountered a dead surface.}
\label{big_cluster}
\end{figure}

\begin{figure}
\centering
\includegraphics[width=0.48\textwidth]{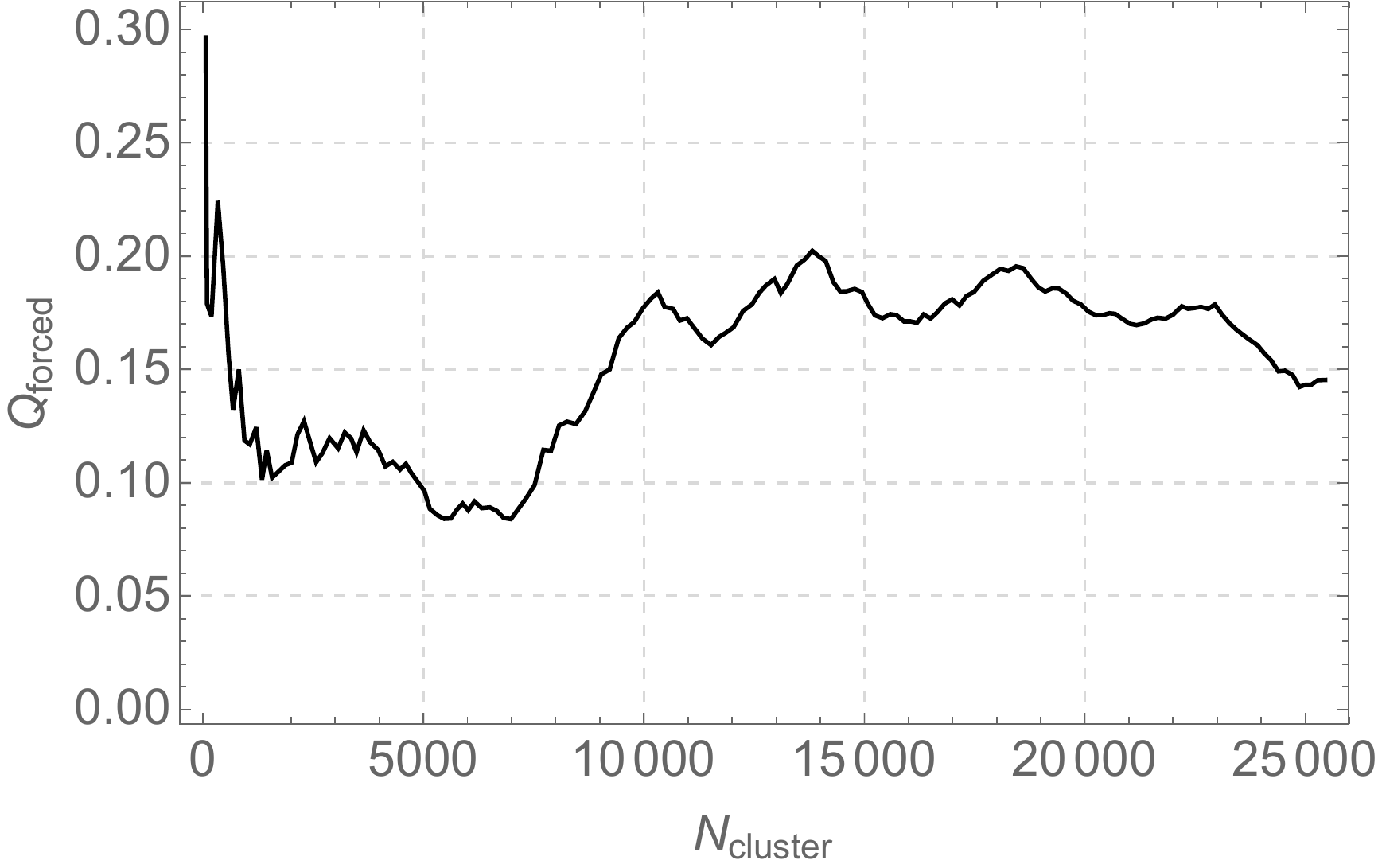}
\caption{Fraction of surface vertices in the cluster of Fig.~\ref{big_cluster} that are forced, plotted as a function of the number of tiles in the cluster during growth.}
\label{fraction_forced}
\end{figure}

An example of simulated growth from the seed of Figure~\ref{net_infinite} is shown in Fig.~\ref{big_cluster}.  It appears that the growth proceeds to infinity, as will be discussed further below.
Figure~\ref{fraction_forced} shows $Q_{\rm forced}$, the number of forced surface vertices divided by the total number of surface vertices, as a function of the total number of tiles in the growing cluster of Fig.~\ref{big_cluster}.  $Q_{\rm forced}$ does {\em not} show dips to very low values that would be associated with growth spurts between nearly completely unforced surfaces.

\begin{figure}
\centering
\includegraphics[width=0.48\textwidth]{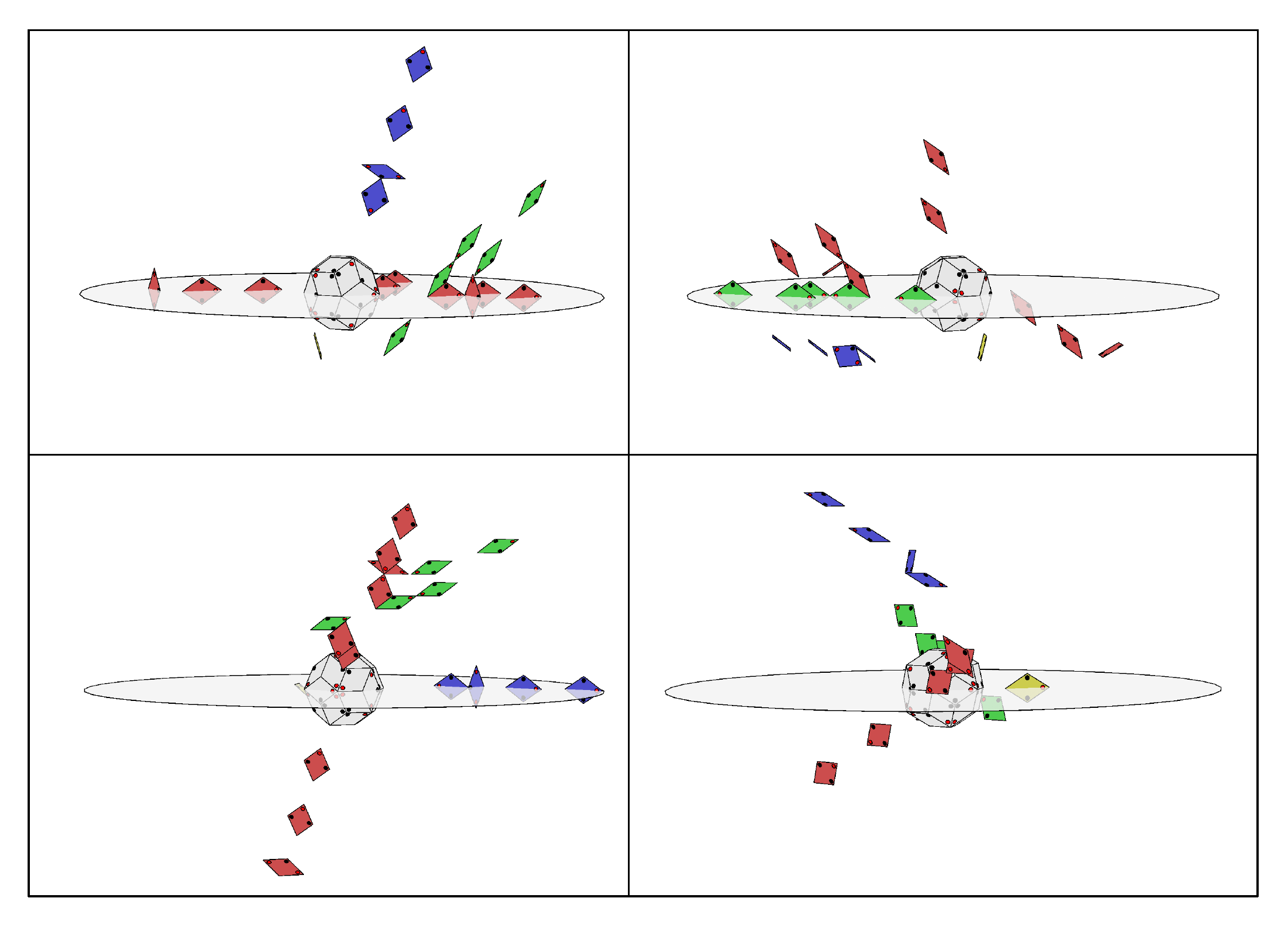}
\caption{Defects confined to worm planes. Here all defective faces are shown from a cluster of approximately 1000 tiles grown from the illegal seed of Figure~\ref{net_infinite}.  The four panels show different views of the same defect structure.  Defect faces in different planes are given different colors. }
\label{defect_faces}
\end{figure}

\begin{figure}
\centering
\includegraphics[width=0.48\textwidth]{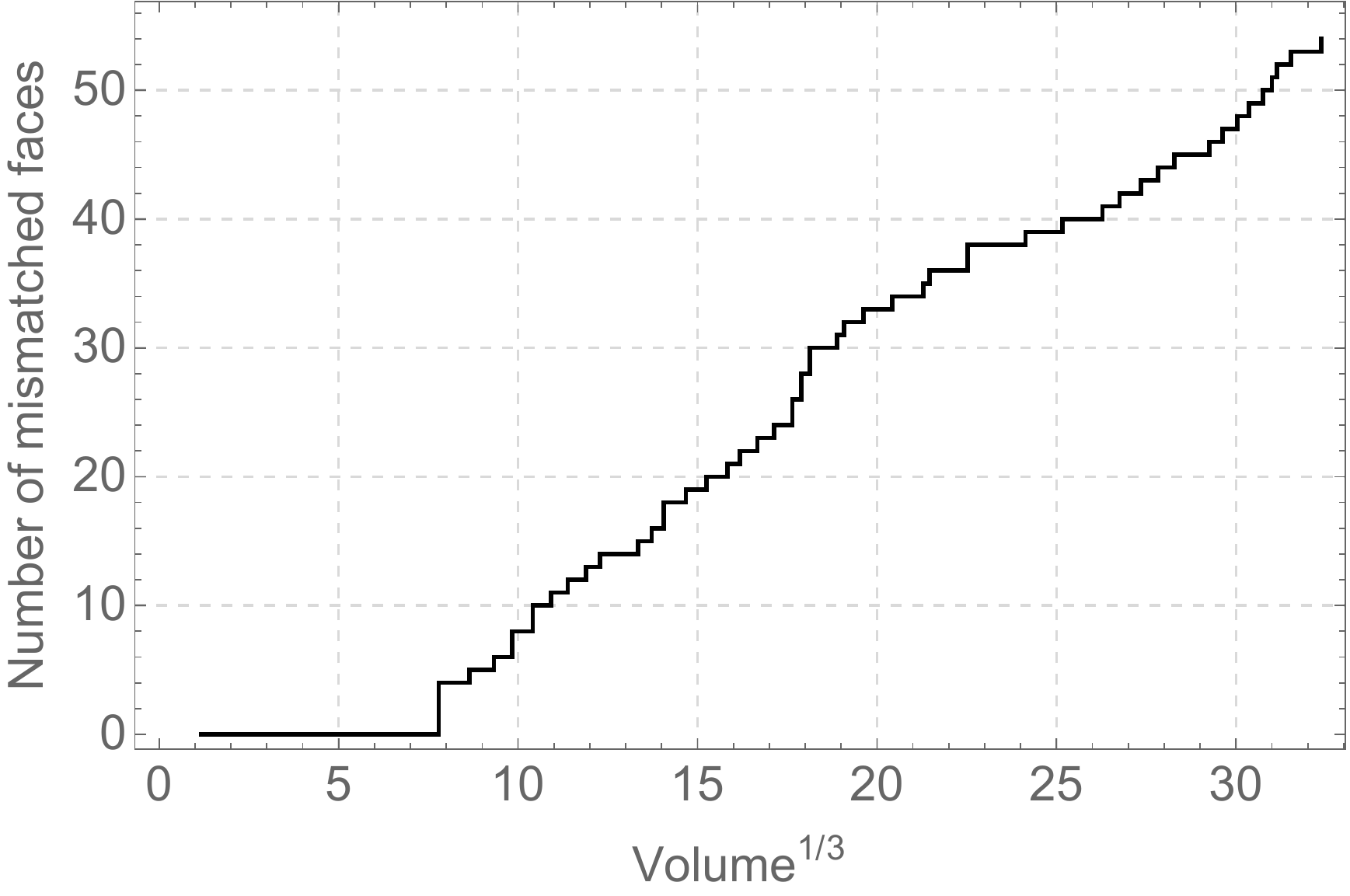}
\caption{Number of defects in the cluster of Fig.~\ref{big_cluster} as a function of cluster volume at the time a given defect appears.  The volume is measured in units of the volume of the oblate tile.}
\label{defect_graph}
\end{figure}

The defects are confined to a subset of the infinite worm planes, as shown in Fig.~\ref{defect_faces}.  In this case, defects appear in the infinite worm planes because opposite faces of the seed specify different orientations for a given plane. As growth proceeds, the given plane is thus divided into two halves of opposite orientation, and a line of defects forms where the two halves meet.  Figure~\ref{defect_graph} shows the number of defects as a function of cluster radius, confirming the expected linear relationship.

\subsection{The probability of infinite growth}
Numerical simulation suggests that the seed of Figure~\ref{net_infinite} does indeed force growth to infinity. Multiple clusters containing several hundred thousand tiles have been grown from this seed, and none has  ever halted due to a lack of forced sites.  We cannot rule out, however, the possibility that growth could be stopped if the order in which forced tiles are added conspires to enclose the seed in a {\em legal} surface.  A defect line separating regions of a worm plane with opposite orientations could bend into loop, leaving an enclosed portion (containing part of the seed) that is flipped but showing no deviations from a perfect worm plane on the surface of the cluster.  If the seed were hidden in this way in a legal cluster, the relevant window for describing further growth would not be constrained by the structure of the seed and would no longer be uniquely specified, so growth would eventually halt for the reasons described above.

The probability of choosing the sequence of forced additions in a way that erases the memory of the seed is clearly quite small and decreases rapidly as the cluster size increases.  If we assume that the cluster grows at a roughly uniform rate in all directions, then the worm plane containing the rays of defects is a growing disk with two points on its circumference marking the points where the defect rays hit the surface. Each time a new layer of tiles is added to the cluster, the defect moves randomly on the boundary by a distance of the order of on tile edge.  Thus the two endpoints of the defect rays executing random walks on the surface with fluctuations that grow as $\sqrt{r}$, where $r$ is the disk radius.  The disk circumference, however, grows at a rate proportional to $r$, making it exponentially improbable that the two walks will meet unless they do so at a very early stage.  We conclude that infinite growth with only infinitesimal phason strain occurs in this model with a probability of order one, where the precise value increases rapidly with the size of the cluster that is taken to be the initial seed.  

The example of infinite growth shown here illustrates a subtle feature of the growth rule.  One might think that a seed could be used for which there are no worm planes with inconsistent orientations on two half planes. The defect in the seed would be encoded in the relative orientations of intersecting worm planes rather than any discrepancies within a single worm plane.  Indeed, the example used here begins with such a seed, as can be seen by the fact that for every pair of opposite faces of the perp-space window both dots appear on the same face.  (When an infinite worm plane is divided into two halves, the interior vertices from the two different halves project onto opposite faces of the window.)  As forced growth proceeds, however, tiles are placed that override the orientations dictated by the original seed, creating half-plane defects.  

\section{Discussion} \label{sec:discussion}
We have shown that the information required to grow a nearly perfect, infinite icosahedral quasicrystal without any backtracking to correct mistakes can be stored in local neighborhoods of the surface sites at all times during the growth.  The growth is nucleated by a small seed and proceeds through the addition of new tiles to randomly selected surface sites, where the probability of attachment is determined by the configuration of existing tiles sharing a single vertex.  The resulting infinite cluster contains a vanishing density of defects; its diffraction pattern would contain the dense set of infinitely sharp Bragg diffraction peaks characteristic of quasicrystals with icosahedral symmetry, and the relative intensity of any diffuse scattering would vanish in the infinite system size limit.  The grown sample can be characterized as a quasicrystal with only infinitesimal phason fluctuations corresponding to inconsistent choices of which faces of the perp-space acceptance window are taken to be closed, but including no points that lie outside the closed window.

We conjecture that the growth of real, rapidly quenched materials is an approximation of the ideal process described here.  The ideal process requires the probability of growth at unforced sites to be strictly zero.  A nonzero probability would lead to occasional additions of the wrong type of tile at a vertex, which would give rise to a finite density of matching rule defects.  There are three classes of such defects.  

The first type of defect involves simple mistakes that create illegal sites where local growth stops until other nearby forced additions promote a correction.  In the second type of defect, two conflicting choices are made for the orientation of a portion of a worm plane, giving rise to additional forced additions that propagate along the worm plane until intersecting worm planes are reached that dictate the proper choice.  In such cases, it is possible that an incorrectly oriented portion of a worm plane will be buried deep below the surface by the time the correct choice is forced.  Defects of this type, which are confined to the interior of a single worm plane, do not disrupt the long range quasiperiodic translational order unless they give rise to the third type of defect, discussed below.  The effects associated with these first two types of defect may be minimized via annealing of a surface layer of finite depth, with the density of the benign intra-worm defects decreasing with increasing depth of the solidification front.   A process of this type appears to have been observed directly by Nagao et al.~in experiments on a decagonal phase~\cite{Edagawa2015}.   We conjecture that similar processes occur in computer simulations of growth mechanisms that involve an advancing solidification front~\cite{Engel2007,Achim2014}.  

The third type of defect presents a greater challenge to the realization of strict quasiperiodic translational order. These arise when the relative orientations of two different parallel worm planes, which may be far apart, conflict with one another.  The defects associated with this type of phason fluctuation become visible as matching rule violations only after additional growth has filled in a bulk region between the two planes.  At that point, the existence of the phason fluctuation is evident, but there is no local indication of where the real problem lies.  Avoiding this type of defect requires the introduction of an illegal seed, which allows forced growth to dictate the proper worm orientations before an incorrectly oriented worm grows too large.  Because defects of this type, which are not identifiable by local tests, do generate phason strains that can disrupt the quasicrystalline order, the growth of a perfect quasicrystal requires a strong separation of scales between the rates of addition at forced and unforced sites.  Further study of the dependence of the size phason strains on the ratio of the two rates and on the depth of the solidification front should provide testable predictions for systems in which the temperatures of the solid and supercooled liquid can be controlled.  The present work shows that the limiting case does allow for essentially perfect growth.

\bibliographystyle{unsrt} 
\clearpage
\normalbaselines 

\newpage
\onecolumngrid

\begin{table}
  \caption{The complete vertex catalog. Each row represents one vertex type, and the 39 rows of the table constitute the entire catalog up to rotations. Within each row, each column represents one face. In a given box, the two numbers specify the icosahedral star vectors (see Eq.~(\ref{eqn:star_vectors})) forming the edges of the face, with overbars denoting negative directions; $a\overline{b}$ indicates that the four vertices of the face are $\boldsymbol{0}$, $\boldsymbol{e}_a$, $-\boldsymbol{e}_b$, and $\boldsymbol{e}_a-\boldsymbol{e}_b$. The arrow indicates the locations of the matching rule dots: an up arrow indicates that a dot is placed near the vertex at the origin, and and a down arrow indicates that a dot is placed near the opposite vertex. The number next to the arrow indicates the location of the second dot: the dot is placed near the vertex located at the tip of the corresponding star vector. Complete tiles can be directly inferred from these faces.}
  \label{tbl:numcat}
  \includegraphics[width=\textwidth]{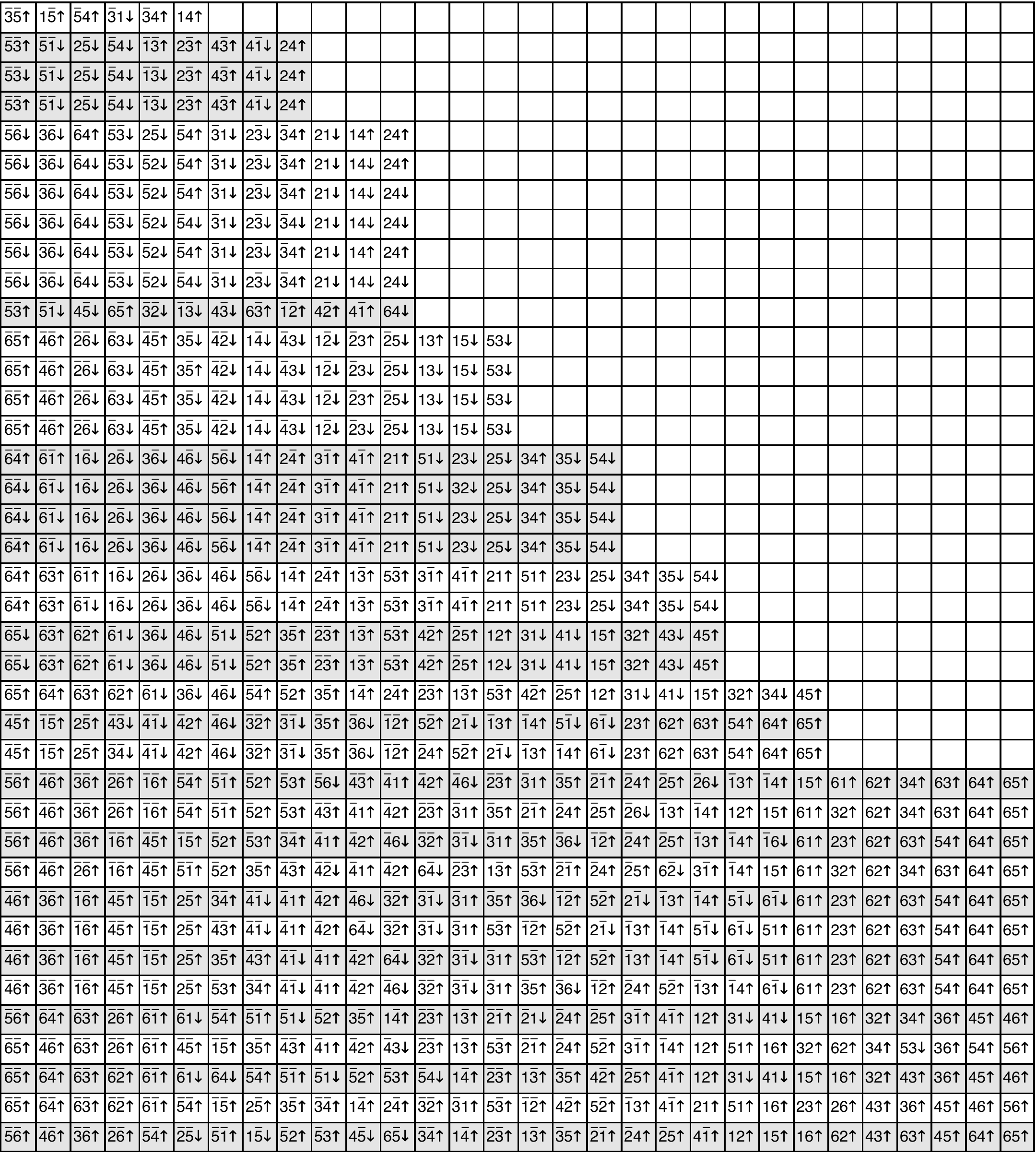}
\end{table}

\end{document}